# The tree reconstruction game: phylogenetic reconstruction using reinforcement learning

Co-authors: Dana Azouri[1,2*], Oz Granit[3*], Michael Alburquerque[2*], Yishay Mansour[3¥], Tal Pupko[2¥], Itay Mayrose[1¥]

[1] School of Plant Sciences and Food Security, Tel Aviv University, Ramat Aviv, Tel-Aviv 69978, Israel.

[2] The Shmunis School of Biomedicine and Cancer Research, Tel Aviv University, Ramat Aviv, Tel-Aviv 69978, Israel.

[3] Balvatnik School of Computer Science, Tel-Aviv University, Ramat Aviv, Tel-Aviv 69978, Israel.

[*] Equal contribution

[¥]Corresponding authors:

Itay Mayrose, itaymay@tauex.tau.ac.il

Tal Pupko, talp@tauex.tau.ac.il

Yishay Mansour, mansour@tauex.tau.ac.il

# Abstract

We propose a reinforcement-learning algorithm to tackle the challenge of reconstructing phylogenetic trees. The search for the tree that best describes the data is algorithmically challenging, thus, all current algorithms for phylogeny reconstruction use various heuristics to make it feasible. In this study, we demonstrate that reinforcement-learning can be used to learn an optimal search strategy, thus providing a novel paradigm for predicting the maximum-likelihood tree. Our proposed method does not require likelihood calculation with every step, nor is it limited to greedy uphill moves in the likelihood space. We demonstrate the use of the developed deep-Q-learning agent on a set of unseen empirical data, namely, on unseen environments defined by nucleotide alignments of up to 20 sequences. Our results show that the likelihood scores of the inferred phylogenies are similar to those obtained from widely-used software. It thus establishes a proof-of-concept that it is beneficial to optimize a sequence of moves in the search-space, rather than optimizing the progress made in every single move only. This suggests that a reinforcement-learning based method provides a promising direction for phylogenetic reconstruction.

# Introduction

A phylogenetic tree is a hypothesis regarding the evolutionary relations among the studied sequences or organisms. Reconstructing a phylogenetic tree for a group of organisms is a fundamental challenge in evolutionary research since Darwin's time. Inferred phylogenies hold a great amount of information regarding the underlying evolutionary process, and their accurate inference is critical for numerous downstream analyses spanning molecular evolution, ecology, and genomics. Leading approaches for phylogeny reconstruction rely on probabilistic evolutionary models that describe the stochastic processes of nucleotide, amino-acid, and codon substitutions[1]. Under the maximum-likelihood paradigm of phylogeny reconstruction, the tree topology, its associated branch lengths, as well as parameters that dictate the substitution rates and the site-specific evolutionary rates are optimized for a given multiple sequence alignment (MSA). Notably, the number of possible tree topologies increases super-exponentially with the number of sequences. When only a few dozen of sequences are analyzed, there are already billions of alternative phylogenetic tree topologies that could potentially describe their evolutionary relationships, rendering the search for the best tree algorithmically challenging.

The computational search for the maximum-likelihood tree topology was previously shown to be NP-hard[2]. Thus, tree-search methodologies rely on a specified heuristic strategy, which must balance accuracy and running time. At present, heuristics employed by the community depend on the intuition of the algorithm developers and their expert knowledge regarding the phylogenetic search space. These are usually based on a hill-climbing rearrangement algorithm that defines neighboring trees[3]. Typically, a search algorithm begins from an initial tree and iteratively replaces the current one via rearrangement to a neighbor with a higher likelihood, until no better neighbor can be found. This procedure results in a tree that is locally better than all its neighbors, but this tree might not be the global optimum. In order to increase the probability of finding the global optimum tree, several techniques have been considered, e.g., initiating the search from multiple starting points, applying simulated annealing (that accepts suboptimal moves with a certain probability)[4], and employing genetic algorithms (in which different areas of the search space are being explored through the use of crossover, mutation and selection operators on a population of candidate solutions)[5]. Our aim here is not to devise a specific novel search strategy, but rather, devise an artificial intelligence (AI) framework, which automatically searches among alternative strategies (policies hereafter) for an optimal one. This framework views the strategy as an evolving entity, which is continuously optimized based on experience, i.e., training data.

Machine learning has been applied to multiple tasks in biology, including molecular-biology, evolutionary, and ecology research[6–15]. Reinforcement learning (RL) is a subfield of machine learning that learns through interaction and is focused on optimizing long-term goals. Over the years RL has had many successes, from playing backgammon (in the 1990's) to playing Go and Attari games (in recent years). RL applications are usually modeled as a Markov decision process, in which an agent (an RL learner) interacts with its environment (the representation of the problem space) in discrete time steps. In each step, the agent chooses an action (a move) to be taken given the current state. The transition between the current state and the next state is influenced by the agent's action, which can be deterministic or stochastic. Another component of RL is a numeric feedback termed the reward, which depends on the current state and on the action taken. Given a sequence of rewards, the return function aggregates multiple rewards to one objective criterion, which implicitly defines the goal of the learner (to maximize the return). Popular returns include the finite horizon return, which considers only the first $H$ returns, for some parameter $H$, and the discounted return, which weights the reward at time $t$ by $\gamma^t$ for some discount factor $\gamma < 1$.

The learner's main task is to select actions that would maximize the return. This is done through a policy, which is a mapping from states to actions, such that given a current state the policy selects an action. For any given policy, a *value function* can be computed, which is the expected return from each state. Our task is to identify the optimal policy, which induces an optimal value function. It is well known that there always exists a deterministic optimal policy. We will consider *terminating environments* in which the interaction terminates eventually. The sequence of states, actions, and rewards from the start state until termination is called an episode.

The RL characteristics of exploratory search and delayed reward, together with the capability of optimizing a sequence of actions (i.e., a policy) for the task of interest, distinguishes RL from other domains of machine learning[16,17]. Accordingly, RL is beneficial in environments where the task is aimed at reaching a winning state at the end of a procedure, rather than aimed at optimizing some myopic (or immediate) reward. When applied to the task of phylogeny inference, RL should allow taking non-greedy steps, which nevertheless, should allow reaching the optimal tree in the fewest number of steps. This is equivalent, for example, to allowing a chess player to make apparently suboptimal moves, such as sacrificing the queen, in order to win the game in the following several moves. Generally, a low (immediate) reward may still lead to an optimal terminal state. Thus, to develop and characterize a full RL algorithm for phylogenetic tree inference, it is necessary to design the algorithm based on a long-term plan by taking into account not only the immediate rewards, but more importantly the future ones.

Setting an RL representation of the phylogenetic tree search dynamics, requires a tailored representation of both the tree topology with branch lengths estimates and the possible actions to be taken, as well as deriving a meaningful immediate reward function. In the phylogeny context, the reward is based on the likelihood change resulting from a local modification of the tree, i.e., the log-likelihood difference between the next and current state (termed hereafter 'likelihood score'). Likewise, a value function considers the entire search path, namely the estimated likelihood scores of subsequent moves.

The RL-based algorithm introduced in this study is based on optimizing a policy for phylogenetic-tree search, with the aim to identify the optimal tree for a given MSA in terms of its likelihood score, relying on previous AI techniques to estimate the likelihood function without actually calculating it[7]. We modeled the phylogenetic tree search problem in a similar way as RL navigation to the highest point on a grid. Defining the grid (environment) as all possible topologies, the state of the agent is the current location, and the action is the move to a new location. The transitions in the environment are deterministic. An optimal policy of an agent would therefore be to reach the topology with the maximum likelihood-score, while taking the minimal number of steps. To account for the length of the path the agent takes till reaching the optimal topology, a discount factor $\gamma$ is set (see Equation 1 in Methods). The discount factor in RL determines the weight the agent assigns to rewards in the distant future relative to those in the immediate future. If $\gamma = 0$, the agent will be completely myopic and only optimize an immediate reward. If $\gamma = 1$, the agent will evaluate each of its actions equally, based on the sum of all its future rewards, thus aiming to reach the optimal configuration but disregarding the number of steps. In the implementation described here, a discount factor $0 < \gamma < 1$ was used as a hyper parameter, which provides an algorithmic incentive to reach the higher likelihood topologies earlier in the trajectory, such that the optimal policy (given a certain $\gamma$ value) corresponds to finding the shortest path from the initial topology to the final topology. The discounted cumulative reward is updated recursively according to Bellman's equation[18] and is estimated using a Q-network[19].

**Figure 1. Modeling phylogenetic-tree search as an RL framework.** A schematic flowchart of the RL framework applied in this study. Given an empirical sequence dataset, the environment represents all phylogenetic trees (states), their possible single-step SPR moves (actions), and the estimated likelihood score. We first extracted feature vectors that represent a state with its actions. These were then fed into the agent's neural network, which outputs a prediction for the best action to be taken in the agent's state, taking into account both immediate and future rewards. The environment was then updated with the reward ($\Delta LL$; the estimated likelihood change) obtained following the action conducted.

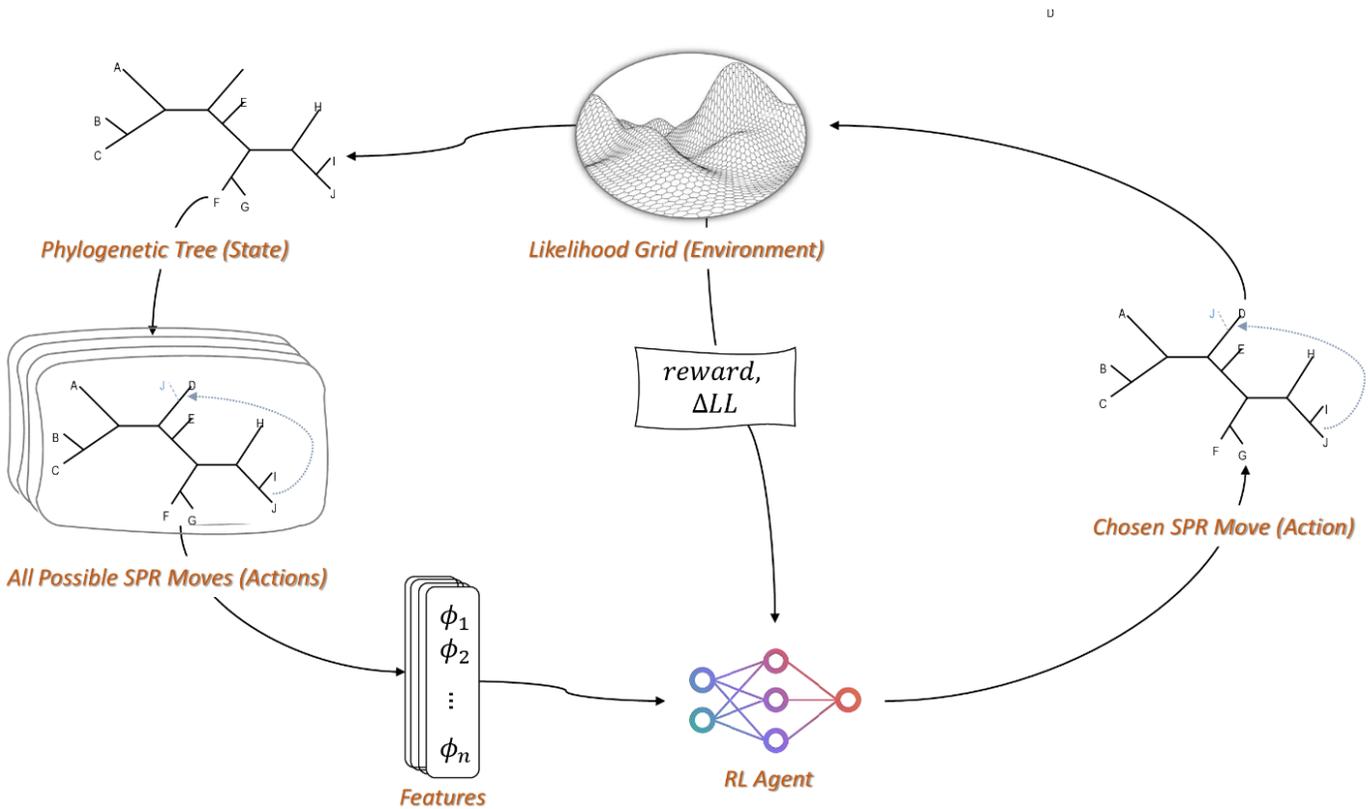

The Q-network, and thus the proposed RL framework, should receive as input a vectorized representation of the states and actions in order to estimate the value function. The representation we designed to suit these requirements is based on a set of tree features that were previously shown to effectively characterize a state-action pair in the context of a likelihood-based phylogeny reconstruction[7]. Specifically, we represented each state-action pair the agent came across during training and testing by calculating 27 features that are based on: the current tree topology, its branch-lengths estimates, and a certain subtree pruning and regrafting (SPR) move[20] to a neighboring phylogenetic tree, by pruning a subtree from the current tree and regrafting it to the remaining tree (see Methods). After each move, i.e., an SPR modification to the current tree, the agent arrives to a new location in the tree space until reaching a predefined end of an episode (see Fig. 1 for a schematic flowchart of the RL framework applied in this study).

The goal of an RL agent is to learn a policy that would make optimal decisions in any given state of the environment. The optimization is performed during a training phase in which an agent plays numerous episodes, allowing it to collect relevant observations (i.e., transitions). That is, by exploring the dynamics of the environment the agent learns the optimal mapping between states and actions for maximizing the long term reward signal[17]. An important issue that needs to be tackled when developing an RL algorithm, as opposed to other types of learning, is how to balance the known trade-off between exploration and exploitation during the training phase. That is, in order to reach beneficial surfaces of high likelihood, the agent has to exploit the good transitions in the tree space it had already experienced. On the same time, it also has to explore unseen transitions, perhaps some that decrease the immediate likelihood gain, in order to make better selection of actions in the future. This was tackled using a known RL technique to sample an action based on its predicted benefit, while allowing some exploration.

Here, we developed an RL strategy for the task of searching for the maximum-likelihood phylogeny. Our method introduces novel approaches for tackling the NP-hard problem of maximum-likelihood tree search by optimizing the exploration strategy itself, which inherently considers suboptimal steps to be taken if they are expected to be beneficial in the long run. Additionally, our method does not require the direct time-consuming calculation of the likelihood function in order to predict an optimal tree. Furthermore, the computational resources needed for using this approach for phylogeny prediction are hardly influenced by the input sequence length. In the following, we first study the potential benefit of looking beyond a single step when using the classic hill-climbing optimization strategy, and demonstrate that taking suboptimal moves can regularly lead to better trees in a subsequent step. Then, a framework based on deep-Q-learning[19] for predicting the optimal tree for a given MSA is introduced. We demonstrate the application of the developed method on a set of unseen data, i.e., on unseen RL environments defined by nucleotide MSAs of up to 20 sequences. Importantly, both training and testing rely on empirical data, which were previously shown to be more challenging for phylogeny reconstruction compared to simulated data[21–24]. Our results show that for this search space, the likelihood scores of the inferred phylogenies are comparable to those obtained from widely-used methods. We then explore the feasibility of applying an agent that was trained on a certain data size on different sizes of the search spaces.

# Results

## The potential benefit of a non-greedy search strategy for phylogenetic reconstruction

The strength of RL lies in its ability to take actions that are suboptimal in the short term for optimizing a long-term reward. To assess the potential benefit of RL in the context of phylogeny-tree search we examined a large number of two-step trajectories and computed the percentage of moves in which choosing two consecutive greedy moves (i.e., as in the greedy approach) would lead to lower likelihood score than choosing a non-greedy move, followed by a greedy one. To this end, for a set of 13,200 starting trees, we generated all possible 1,082,400,000 two-step trajectories. For each starting tree, we located the best tree (i.e., the best two-steps neighbor) in terms of the likelihood score. We then quantified the fraction of starting trees for which the greedy approach was not optimal. This analysis revealed that the greedy approach was suboptimal in 33% and 41% of the cases for datasets of size 7 and 12, respectively. Interestingly, some of the intermediate moves that led to trees with higher likelihood than the greedy approach were among the worst possible first moves (Fig. 2). Although the analysis was not prolonged for more than two steps ahead, this result implies that the strict stepwise greedy optimization is not necessarily the best strategy to traverse the tree topology space, even when the search space is rather limited.

**Figure 2. The ranking percentiles of beneficial suboptimal first moves.** The distribution of the percentiles of first moves that led to better trees than two-step greedy moves. The box inside each violin shows the quartiles of the dataset with the white dot being the median, while the whiskers extend to show the 1.5 × IQR past the low and high quartiles.

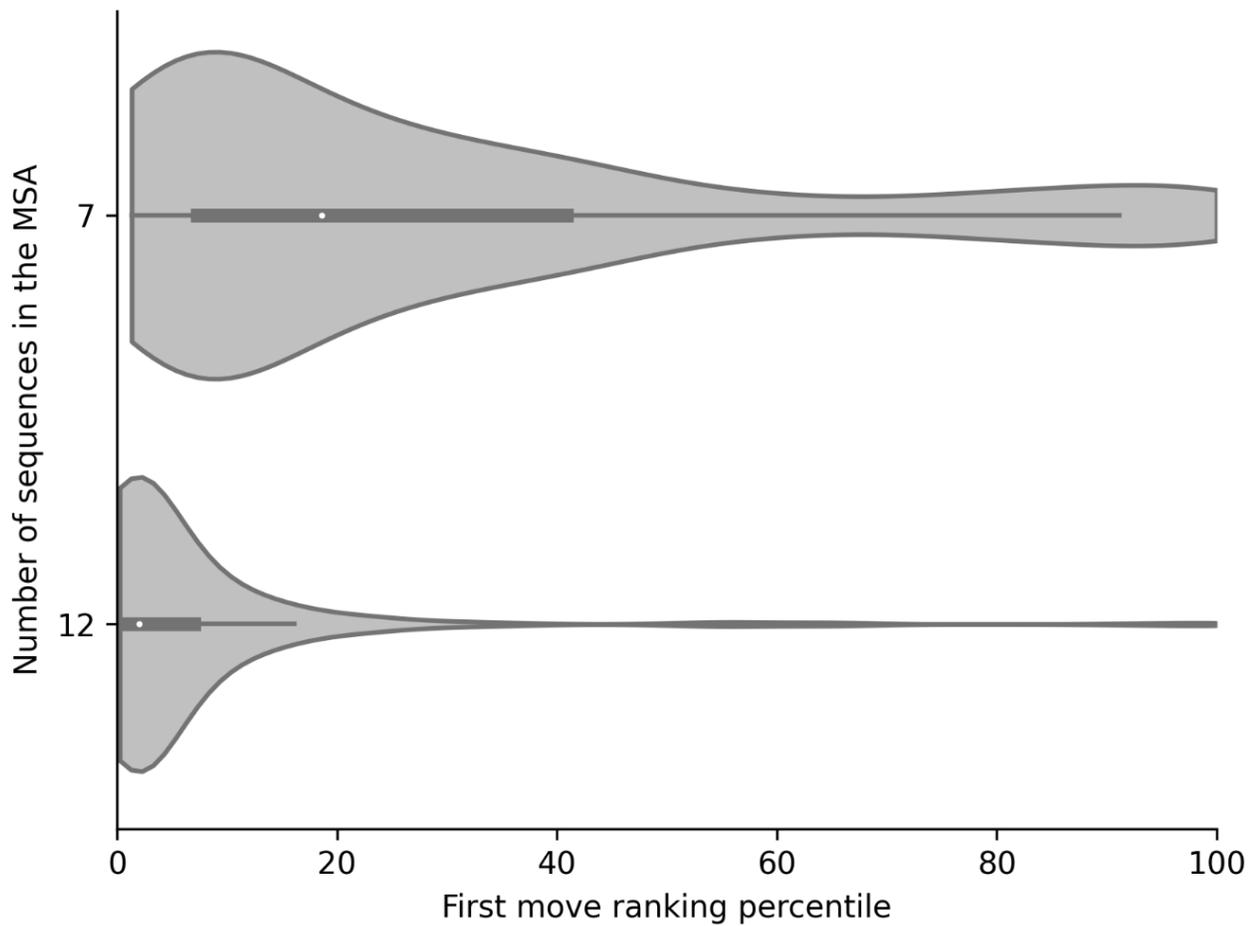

## Performance evaluation of the proposed RL framework

We developed a tree-search framework that is entirely based on RL, and tested its performance. For training the RL model, we assembled a large collection of transition data from a database of empirical MSAs[7] (see Methods). We defined transition data as all the observed shifts from a certain state-action pair to an adjacent state-action pair, together with the corresponding likelihood estimates. We first focused on relatively small datasets, containing at most 12 sequences (i.e., a space size of up to ca. $10^9$ topologies). For these datasets, for each state the agent came across, we explored all possible immediate SPR moves during training and testing. For larger datasets that contained 15 and 20 sequences (i.e., a space size of up to ca. $10^{20}$ topologies), we restricted the range of possible actions from each state in order to make the training of agents feasible for the scope of this proof-of-concept study (see Methods).

The RL algorithm aims to optimize the entire search path from the starting tree to the global maximum. Therefore, throughout this study we measured the agent's performance at the end of an episode according to the improvement the agent achieved relative to improvement obtained by RaxML-NG[25] from the same starting tree (see Methods). Thus, an agent that achieved the maximal observed improvement received a score of 1, while an agent that achieved an improvement of 150 likelihood points relative to the starting tree, but 50 likelihood points less than the estimated global maximum, received a score of 0.75.

Typical examples of the likelihood improvement as the agent progresses in the search space is shown in Fig. 3. In these examples, we compared the trajectory of a trained agent to a hill-climbing fully-greedy strategy (i.e., evaluating the log-likelihood of all possible neighbors), and to the maximum-likelihood score obtained by running RaxML-NG (i.e., the final likelihood only), when all three searches were initiated from the same random tree. Three different examples are presented, where: (a) the RL agent did not reach the best-known tree; (b) the RL agent discovered the optimal tree, while taking fewer moves than the fully-greedy procedure (five compared to six moves) by taking suboptimal moves; (c) the RL agent converged to a better tree than the greedy search.

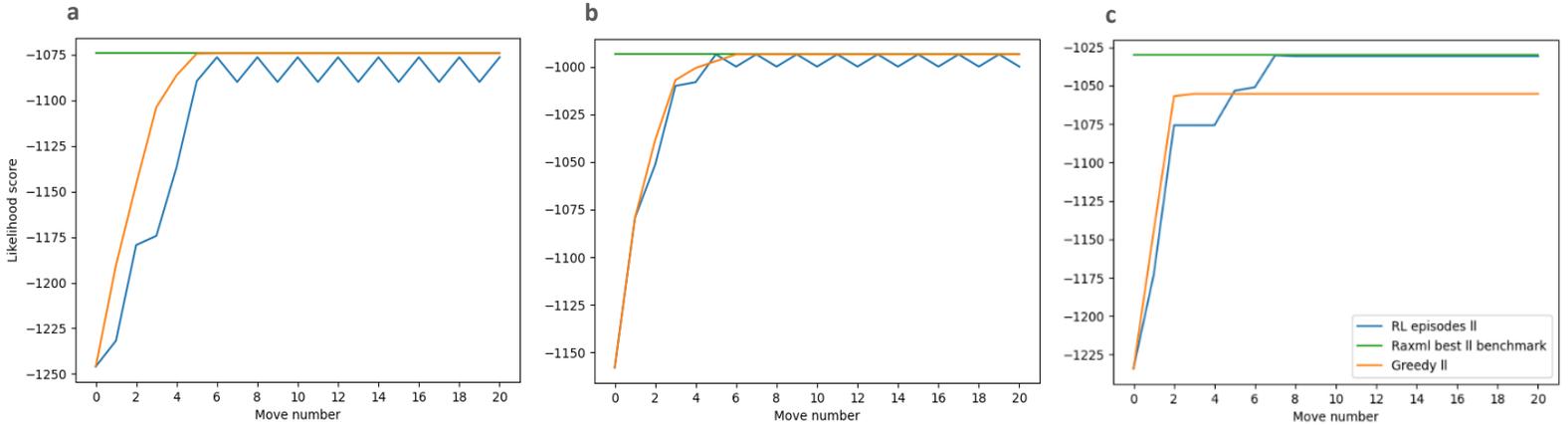

**Figure 3. Typical examples of the log-likelihood improvement as the search progresses.** An example of the likelihood gain of a trained agent (blue), a hill-climbing-fully-greedy strategy trajectory (orange), and a maximum-likelihood tree obtained by RaxML-NG (green). Different panels represent different tests, on datasets containing 12 sequences (a,b) and 15 sequences (c).

*Accuracy for datasets of relatively small size*

We evaluated the performance of the trained RL model on unseen test data. First, we examined data composed of seven sequences. For this challenge of searching in a space of size $10^3$, both the heuristic software (i.e., RaxML-NG) and our RL model converged to the optimal tree with an average accuracy of 0.99999 (95% confidence interval of 0.999 - 1). We next evaluated the performance on a much larger search space, such as that defined by MSAs containing 12 sequences (i.e., search space of size 654,729,075 topologies). The average accuracy score of this trained model was 0.969 (95% confidence interval of 0.945 - 0.993). This indicates that the trained RL agent successfully learns a search strategy that can be well generalized for empirical datasets of various sources.

The above results were obtained with 10 datasets for generating the training observations and 2,000 training episodes. These values were selected by analyzing the dependence between the prediction accuracy on the validation set and the number of MSAs used to generate the training data, focusing on datasets with 12 sequences. To this end, we increased the number of different empirical datasets based on which we generated the training observations from 1 to 10, 20, and 30 (but keeping the total number of episodes and transitions constant) and compared the performances (Supplementary Fig. 1). This analysis indicated that using only a single dataset for learning is significantly inferior to all other sizes (P-value < 0.03 for one-way ANOVA test for the means), but using more than 10 datasets does not significantly improve the performance (P-value > 0.64 for one-way ANOVA test for the means when comparing 10 to 20 and 30 datasets). Consequently, 10 different empirical datasets were used to collect the training data. To further investigate the main factors affecting the performance of the RL agent, we sought to investigate the impact of the number of episodes in the training phase on the validation accuracy. The accuracy increased as a function of the number of episodes (P-value < 0.004; Pearson correlation coefficient for testing non-correlation of the means), reaching a plateau at around 2,000 episodes. Though the increase in accuracy was statistically non-significant when increasing the number of episodes in the range between 1,500 to 5,000 (Supplementary Fig. 2; P-value > 0.37 for one-way ANOVA test for the means), the best accuracy was obtained when 2,000 episodes were used during test. To balance runtime and accuracy, the results across the entire analyses are presented using 2,000 episodes and 10 distinct empirical MSAs to generate the training data.

*RL for large search space*

Search spaces of datasets containing 15 and 20 sequences are of size $7.9 \times 10^{12}$ and $2.2 \times 10^{20}$ topologies, respectively. For these datasets, features extraction of all possible neighbors of a given tree, either at the learning stage or when searching for the best tree, is computationally demanding. Thus, we limited the number of considered neighbors of a given state by applying a restriction on the SPR moves, considering only local changes in the tree topology as commonly performed in various tree search heuristics[7,26,27] (see Methods). When applying this procedure both in training and in testing, the average performance of the trained models for 15 sequences was 0.999 (95% confidence interval of 0.998 - 1.001). This suggests that narrowing the range of possible neighbors should be considered as a technique for training RL agents and inferring the phylogenies for datasets with large phylogenetic search spaces. When datasets with 20 sequences were considered, the test accuracy was lower, i.e., the average test performance was 0.89. We speculate that this performance could be improved using alternative, more exhaustive, data collection methodologies (see Discussion).

*Employing pre-trained agents across data sizes*

Our learning so far concentrated on RL training on datasets of specified size. To assess the potential of using agents that were trained on a specific dataset size to solve the phylogenetic search problem for varied number of sequences, we sought to apply zero-shot testing[28]. Specifically, we investigated the predictive power when testing pre-trained agents of up to 20 sequences on datasets with fewer sequences and found comparable performance (Table 1). For example, the performance of a zero-shot agent trained on datasets containing 15 sequences on unseen environments of datasets containing 12 sequences obtained an averaged accuracy score of 0.973, which is slightly better than that obtained by an agent that was trained and tested on an environment of 12 sequences (average accuracy of 0.969). Overall, this analysis indicates that a transfer between environments of different sizes does exist and that this approach could potentially assist in solving varied phylogenetic-search space environments.

**Table 1. Accuracy scores of zero-shot experiments.** *The table details the performance of each pre-trained agent of a certain dataset size (row) to each other smaller dataset size (column). Each cell shows the accuracy score of the trained model, averaged over the test datasets.*

|    | 7     | 12    | 15    | 20    |
|----|-------|-------|-------|-------|
| 7  | *0.999* | -     | -     | -     |
| 12 | 0.999 | *0.969* | -     | -     |
| 15 | 0.998 | 0.973 | *0.993* | -     |
| 20 | 0.999 | 0.93  | 0.993 | *0.892* |

*Running times*

In this study, we focused on developing the conceptual aspects of RL-phylogenetics, and as part of this we developed a prototype implementation. This prototype did not undergo cycles of optimization, e.g., a large portion of the computational runtime is devoted to feature extraction, which in the current version is implemented inefficiently in Python. For 15 sequences, for example, the training of an RL agent took 600 CPU hours. However, once the agent is trained, the time required to predict the optimal tree takes a few seconds. Specifically, we compared the runtime required to reconstruct the optimal tree to that of RaxML-NG. For datasets with 15 sequences running the trained agent took 8.7 s on average (ranged between 8.4 and 9.3 seconds), of which, 7 s for extracting the features, and 1.7 s for all other computational tasks, e.g., estimating the $Q$ function. Noticeably, the running time does not depend on MSA length (Fig. 4). For the same datasets, the likelihood computation of RaxML-NG took 8.7 s on average, but these varied widely from less than half a second for short MSAs (up to 800 base pairs) to 18 s for very long ones (more than 16,500 base pairs). The same trend was observed when datasets with 20 sequences were considered (Fig. 4).

**Figure 4. Running time.** The average inference running time in seconds (y axis) relative to the length of the sequences analyzed (x axis; 100 data points binned to 17 groups). In blue and orange are the average running times of inferring the optimal tree for datasets with 15 sequences using the RL trained agent and RaxML-NG (with the same single-random-starting point), respectively. Similarly, in green and red are the running times for datasets containing 20 sequences, of the RL agent and RaXML-NG, respectively.

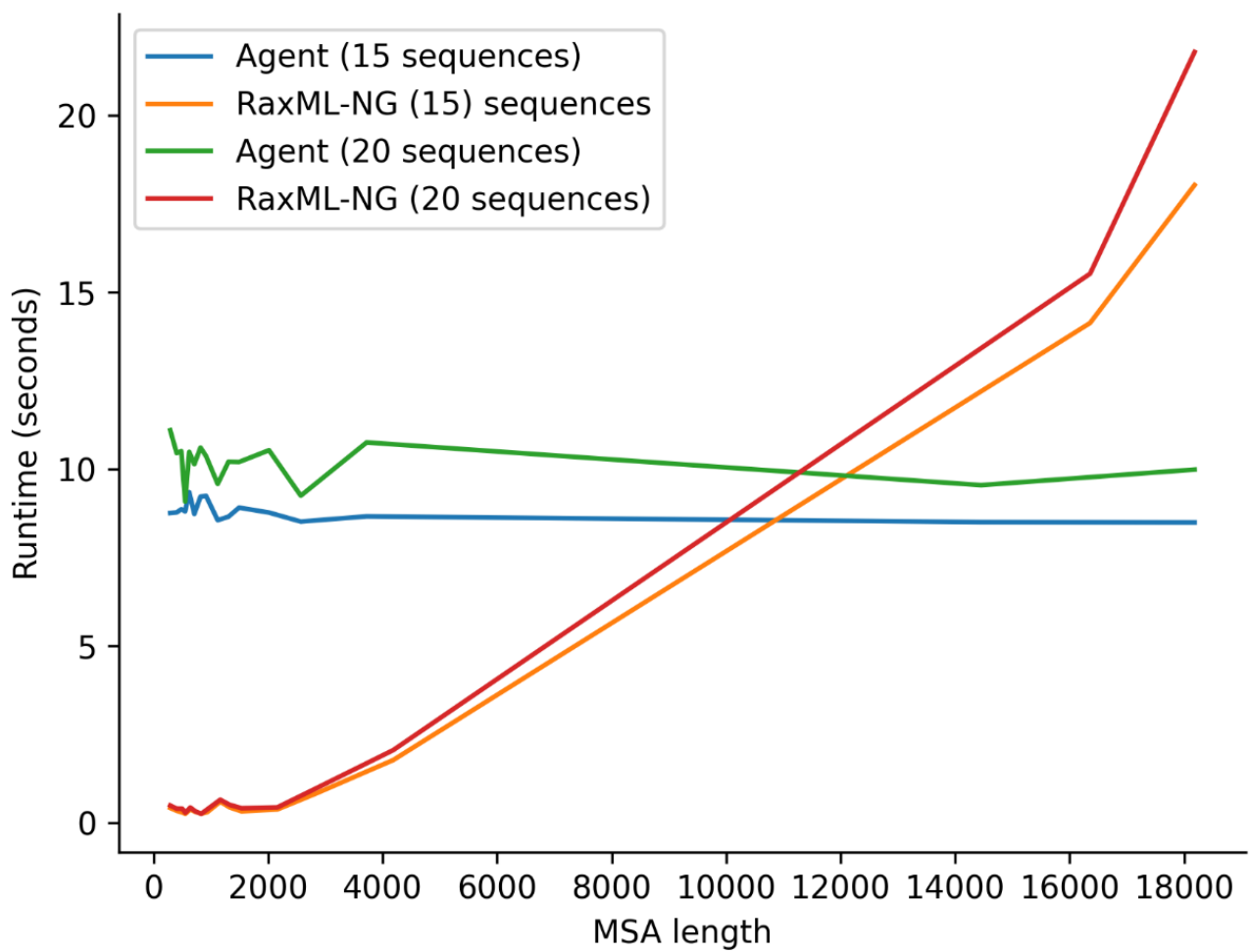

# Discussion

For many biological domains, spanning diverse fields such as ecology, genomics, systematics, and epidemiology research, an accurate inference of the underlying phylogeny is indispensable. As such, the development of more accurate phylogeny-reconstruction techniques is an ongoing effort that continuously progressed with the type and size of data analyzed, the computational resources available, and algorithmic developments. Numerous computational techniques were imported from the fields of statistics and computer science to improve phylogenetic tree reconstruction. These include treating character evolution as a Markov process[29], Branch-and-Bound[30], Markov chain Monte-Carlo[31], genetic algorithms[5], simulated annealing[4], and more recently, machine learning[7,11,12]. Despite these improvements, commonly used algorithms still lack the ability to provide an optimal solution. In this study, we propose an out-of-the-box AI approach for phylogenetic reconstruction, namely, reinforcement learning.

The idea of introducing RL algorithms to the task of finding the optimal phylogenetic tree is based on the concept of optimizing a strategy for the tree search, rather than incrementally optimizing the likelihood gain within a series of steps. RL includes several aspects that together could prove particularly beneficial to phylogeny inference. First, similar to simulated annealing, it allows taking suboptimal steps as part of the search strategy. Our results above demonstrate that this often enables more efficient convergence to optimal trees. Second, and unlike any other existing approach, our algorithm directly optimizes a policy based on empirical training data, without the need of predetermined heuristics. This means that an agent can decide to be greedy or to take suboptimal moves according to the specific characteristics of the data and the specific position in the tree space. Third, our agent moves without optimizing the likelihood directly, potentially reducing running time, especially for long sequences.

When a phylogenetic tree is provided as input to machine learning algorithms, it must be represented as a vector. In a recent study we represented a tree and its SPR neighbors as a vector of 19 features, and showed that we could predict optimal SPR moves without computing the likelihood function[7]. In this work, we exploited such tree representation technique for training an RL agent, which can successfully traverse previously unseen phylogenetic spaces of empirical datasets. This study could thus serve as a benchmark for different representations of a phylogenetic search space, which reportedly until current days, was missing. We expect that additional improvement in representing trees and alignments would further improve RL-based tree search.

Two recent studies employed RL to phylogeny[32,33]. In these studies RL was used in the context of distance-based methods, which are known to be faster, albeit less accurate than likelihood-based methods[34,35]. These studies and ours showed the potential of using RL for the complicated tree search problem and emphasized the challenge of training an agent on topological spaces of more than 20 sequences. While our algorithm and representation are not theoretically limited by the number of sequences in the data, at present, accuracy was not satisfactory for datasets with more than 20 sequences. We believe that promising future directions towards the application of RL to large datasets should concentrate on the following: (1) Improving the training data. This includes the size of the training observations, as well as its quality. Increasing the number of training examples necessitates training the agent longer, which depends on the availability of computational resources. In this regard, transfer learning should enable repeatedly using previous models trained on small datasets as the starting point for learning larger ones[36]. As for the data quality, it requires developing means to collect training observations of those cases that would maximize the learning of the agent. For example, by collecting more observations from regions of high likelihood, which could provide valuable information for traversing these important parts of the likelihood surface; (2) Improving code efficiency. Although the running time of the proposed methodology is hardly affected by the input sequences lengths and thus is suitable for large-scale data, a more efficient implementation with regard to feature extraction could enable better usage of the computing resources available, particularly during training; (3) Using an alternative, automatic, representation of the tree search space. For example, it has been recently proposed to represent tree topologies with embedded node features based on graph neural networks[37]. This direction of extracting learnable topological features can potentially better capture the complexity of empirical phylogenetic environments, without requiring to hand-craft additional features.

Another possible direction for improving the effectiveness of RL for phylogenetics could be considering alternative immediate reward functions, e.g., directly calculating the likelihood-function as the reward during inference instead of estimating the likelihood change. Additionally, while in this work we considered SPR actions only, the combination of complementary neighborhood definitions for local-search phylogenetic-reconstruction algorithms, such as nearest-neighbor interchange (NNI)[38] and tree bisection and regrafting (TBR)[39], could be considered when modeling the tree search dynamics. Expanding the range of possible actions could thus help the agent fine tune the search strategy when it is in low or high likelihood regions. Lastly, there are a large number of variants of RL algorithms. As part of the development of the current implementation, we have examined the applicability of alternative RL-based schemes, e.g., policy-based algorithms. This procedure demands more computation resources and was attempted for relatively small data sizes of up to 12 sequences. However, other existing RL frameworks could prove beneficial for the task of phylogenetic reconstruction.

The main conceptual novelty of our approach is to view phylogenetic tree reconstruction as a dynamic game, in which the rules are specified, but the winning strategy is unknown and difficult to optimize. In such a case, better inference is obtained following numerous games generated in-silico. We expect that, with time, RL will be introduced for additional evolutionary genomics optimization problems, including multiple sequence alignment, synteny inference, and elucidating complex patterns of population dynamics.

# Methods

## A reinforcement-learning algorithm for predicting the maximum-likelihood phylogeny

*The environment*

We defined the environment using $(S, A, R)$, where $S$ is the state space, i.e., all possible trees given a set of aligned sequences, and $A$ is the set of possible actions, i.e., all possible SPR moves given a tree topology. $R$ is the immediate reward function following a transition from state $s$ to $s'$. In our setting, $(s, a)$ deterministically determines the next state $s'$. $R(s, s')$ is defined as the log-likelihood difference, scaled by $LL_{NJ}$ (the log-likelihood of the reconstructed BioNJ[40] tree as implemented in PhyML 3.0[41]) so that the reward function would have the same magnitude across different datasets: $R(s, s') = \frac{LL_{s'} - LL_s}{LL_{NJ}}$.

### The Features

Each state-action pair $(s, a)$ is represented by a set of 27 phylogenetically informative features from an input tree (Table 2). The feature vector, $\phi(s, a)$, captures properties of the current state (the topology and its branch lengths) and the action (one possible SPR move). Of these, 19 were previously developed in the context of predicting the optimal neighbor as part of a tree search[7] and capture, for example, features related to the topological differences between the starting and resulting trees and properties related to their branch lengths. Eight additional features were implemented in this work and are based on non-parametric bootstrap computations[7].

*Table 2. The features used to represent a state-action pair. The table lists the 27 features on which the RL state-action representation is based. Features 20-23 were extracted base on the UPGMA algorithm[142], while features 24-27 were based on the NJ algorithm[107].*

| # Feature | Feature name | Details | Represented action | Tree considered |
|---|---|---|---|---|
| 1-19 | Detailed in Table 1 in Azouri et al.[2] | | | |
| 20 | Bootstrap – UPGMA | The approximated bootstrap support of the internal branch (that defined a split) that was being pruned or regrafted. | Pruning | Current tree (s) |
| 21 | | | Regrafting | |
| 22 | | More specifically, 10,000 bootstrapped trees were generated based on both UPGMA and NJ distance matrices, as implemented in Biopython package[143] version 1.79. Splits that do not exist in those 10,000 trees received a value of zero, while splits that lead to an external node received a value of 100. | Pruning | Next tree (s'), namely the resulting tree following an SPR move |
| 23 | | | Regrafting | |
| 24 | Bootstrap – NJ | | Pruning | Current tree (s) |
| 25 | | | Regrafting | |
| 26 | | | Pruning | Next tree (s'), namely the resulting tree following an SPR move |
| 27 | | | Regrafting | |

*The algorithm*

The developed RL algorithm is based on a Deep Q-network (DQN)[19], a model-free and off-policy RL algorithm. In the DQN setting, the agent learns a value function, named the quality function $Q(s,a)$, which represents the estimated benefit of a specified action in gaining some future reward, given a certain state. More specifically, we implemented a neural-network $Q_\theta$ (with weight parameters $\theta$) that predicts the quality function of a state-action pair, given the feature vector $\phi(s,a)$. This predicted value is termed $Q(s,a)$ and is explained in more details below. Starting from state $s$ we estimated $Q(s,a)$ for all possible SPR actions and chose the action with maximal $Q(s,a)$, which defines the next state (Algorithm 1). The number of unique state-action pairs to be computed when conducting a move is $2(n-3)(2n-7) = O(n^2)$ where $n$ is the number of sequences in the input MSA[39]. The starting state for each episode, $s_0$, was randomly sampled (using RaxML-NG[25] random tree generator), such that the agent could start the trajectory from anywhere in the tree space.

---

**Algorithm 1** Deep Q-Learning of Tree Reconstruction - Inference

Given a Feature-Extractor function $\phi$ and an action-value function $Q_\theta$
Sample starting tree $s_0$

**for** $h = 0...H$ **do**
    **for** all possible SPR-moves $a$ in $s_h$ **do**
        Evaluate $Q_\theta(\phi(s_h, a))$
    **end for**

    select $a_h = \arg\max_a Q_\theta(\phi(s_h, a))$
    Execute SPR-move $a_h$ and observe next tree $s_{h+1}$

**end for**

**return** $s_h$

---

*The model*

The main strength of Q-learning lies in its ability to construct a policy that maximizes the commulative reward[17]. In deep Q-learning, neural networks are trained to estimate the value of the $Q$ function for unseen states and thus it combines Q-learning with a deep artificial neural network (ANN). The recursive form of the optimal return function, known as the Bellman equation[18] is:

(1) $$Q^*(\phi(s_t, a_t)) = r_t + \gamma \max_a Q^*(\phi(s_{t+1}, a))$$

Where $Q^*$ is the optimal state-action value function, and $r_t$ is the immediate reward obtained at time-step $t$. The $\gamma$ hyperparameter is the discount-rate, a constant $0 \leq \gamma \leq 1$, which weights rewards from the uncertain far future less than the ones in the fairly confident near future. That is, a reward received $k$ time steps in the future is worth only $\gamma^{k-1}$ times what it would be worth if it were received immediately. Of note, there exists a $\gamma = 1 - \epsilon$, $\epsilon > 0$, for which the optimal policy corresponds to the shortest path from the starting topology to the topology with the highest likelihood (up to $H$ SPR moves away).

We would like the agent to learn a policy that does not involve taking an unlimited number of actions, i.e., we would like to balance the runtime with the improvement in likelihood. This balance is controlled by the horizon hyperparameter, denoted as $H$, which specifies the number of actions taken from the starting tree (Algorithm 2). Importantly, the specific $\gamma$ value inspires a different optimal policy (see Supplementary note 1). Of note, each search stops after a predefined number of $H$ steps. This is termed an episode. As stated above, the total number of episodes during learning is also a hyperparameter. Following each episode, the network weights are updated. Specifically, as in standard DQNs, we used the experience replay method to hold the agent's training trajectories, i.e., a buffer of a pre-determined size containing transition observations (state-action pairs together with their rewards and next state-actions). At the end of each episode, $H$ new memories are added to the buffer (and the $H$ oldest memories are discarded), and the ANN is trained based on a batch of trajectories sampled (with replacement) from the memory buffer 'time-to-learn' times (Algorithm 2). The sizes of the memory buffer and the batch, as well as the 'time-to-learn', are hyperparameters of the algorithm.

Additional hyperparameters are related to the deep network architecture and the learning dynamics. These include the number of fully-connected hidden layers, the number of neurons in each layer, the activation function, the loss function, the optimization algorithm, and the learning rate (Supplementary Table 1).

To control the exploration-exploitation tradeoff during training we allowed the agent to take suboptimal moves with respect to the $Q$ function. We used the SoftMax exploration strategy that selects an action $a$ based on the following probability: $\frac{e^{Q(s,a)/T}}{\sum_{a'} e^{Q(s,a')}}$. This allows greater value actions to be selected with greater chance, yet permitting some randomness. $T$ is a hyperparameter that controls the level of exploration.

We implemented the ANN in Python using PyTorch[42]. The above hyperparameters were optimized via Optuna framework[43], which is an automatic hyperparameter optimization package, particularly designed for machine learning (summary of the model hyperparameters values and further details are described in Supplementary Table 1).

## Algorithm 2 Deep Q-Learning of Tree Reconstruction - Training

Given a Feature-Extractor function $\phi$
Initialize action-value function $Q_\theta$ with a random set of weights $\theta$
Initialize replay buffer $\mathcal{D}$ to capacity $N$

**for** $episode = 1...M$ **do**

    Sample starting tree $s_0$
    **for** $h = 0...H$ **do**

        **for** all possible SPR-moves $a$ in $s_h$ **do**
            Evaluate $Q_\theta(\phi(s_h, a))$
        **end for**

        **if** $h > 0$ **then**
            $a' = \arg\max_a Q_\theta(\phi(s_h, a))$
            Store the transition $(\phi(s_{h-1}, a_{h-1}), r_{h-1}, \phi(s_h, a'))$ in $\mathcal{D}$
        **end if**

        Sample from exploration policy $\pi$: $a_h = \pi(s_h)$
        Execute SPR-move $a_h$, collect reward $r_h$ and observe next tree $s_{h+1}$
    **end for**

    **for** $t = 0...times-to-learn$ **do**

        Sample mini-batch of transitions $(\phi(s_j, a_j), r_j, \phi(s_{j+1}, a'_j))$ from $\mathcal{D}$
        Perform a gradient descent step using
        $Q_\theta(\phi(s_j, a_j)) = r_j + \gamma Q_\theta(\phi(s_{j+1}, a'_j))$
        with respect to the online parameters $\theta$
    **end for**

**end for**

*The performance metric*

To measure the performance at the end of an episode, we computed the following metric: let $LL_{gain}$ be the log-likelihood difference between the final tree and the starting tree. Let $LL_{rax}$ be the log-likelihood difference between the maximum-likelihood tree and the starting tree, as obtained by executing RAxML-NG from the same starting tree. Both resulting topologies were subject to branch-lengths optimization. The ratio between these two terms ($LL_{gain}/LL_{rax}$) is a number that reflects the improvement in likelihood score achieved by an agent, and can be used to compare the performance between different datasets.

## Empirical data preparation

We selected all empirical datasets with 7, 12, 15, and 20 sequences from the training data collected in Azouri et al.[7]. These represent nucleotide coding alignments[44], user-submitted phylogenies from TreeBase[45], plant phylogenies reconstructed using the OneTwoTree pipeline[46], and genomic sequences that were aligned according to the tertiary structure of its encoded proteins[47]. For each dataset size (i.e., the number of sequences in the alignment), 30 MSAs were randomly fixed as validation data, 10 MSAs were randomly fixed as test data, and from the rest, 10 datasets (unless otherwise specified) were randomly sampled to generate the training samples.

*Transition data collection*

To apply the memory buffer method, the transition observations (state-action pairs together with their corresponding likelihood estimates, and the corresponding next state-action pair) need to be collected through many training episodes. To this end, each episode was initiated from a random tree and the obtained reward was calculated (using RaxML-NG) and stored for each transition taken. Precisely, a model that is trained for 2,500 episodes of 20 SPR moves each is essentially trained over $25{,}000 \times 20 = 50{,}000$ training observations. The substitution rate parameters of a GTR+I+G model[21] were optimized once for each dataset, based on a reconstructed BioNJ[40] tree as implemented in PhyML 3.0[41] and were then fixed for the following likelihood calculations of the respective dataset.

When datasets of size 15 and 20 sequences were considered, the computational resource required to compute the features for all neighbors were beyond the scope of the study conducted here. Therefore, we limited the space of possible neighbors by applying a restriction on the range of SPR moves, allowing each pruned subtree to be regrafted up to a pre-defined radius. This radius defines the number of branches in the path between the pruned and regrafted branches, not including these branches (Supplementary Figure 3). Setting a radius of four narrowed the neighborhood space to a feasible task. Additionally, when datasets with 20 sequences were considered, we applied an alternative approach to collect experiences, which proved superior to the one used for smaller data sizes (see Supplementary note 2).

*Data collection for the two-step pre-analysis*

We collected from the training data all MSAs containing 7 and 12 sequences (i.e., 51 and 81 datasets, respectively). Next, 100 random starting trees were reconstructed for each dataset using RAxML-NG. We then obtained all their respective 32,640,000 and 1,049,760,000 possible two-step trajectories. Precisely, we: (1) obtained all single-step SPR neighbors for each starting tree and recorded all likelihoods; (2) recorded all likelihood scores of the single-step SPR neighbors of each of the latter trees. This allowed us to identify the best two-step neighbor of each starting tree, as well as the best neighbor reached by applying the single-greedy step twice sequentially. The likelihoods throughout this analysis were computed using RAxML-NG, allowing for branch-lengths optimization. The substitution rate parameters were optimized once for each dataset, based on a reconstructed BioNJ[40] tree as implemented in PhyML 3.0[41] and were then fixed for the following likelihood calculations of the respective dataset, assuming the GTR+I+G model[21].

# Data availability

The datasets contained within the empirical set have been deposited in GitHub with the identifier https://github.com/michaelalb/ThePhylogeneticGame[48].

# Code availability

The code that supports the findings of this study was written in python version 3.9.7 has been deposited in GitHub with the identifier https://github.com/michaelalb/ThePhylogeneticGame[48].

Computation of likelihoods were executed using the following application versions: PhyML 3.0[41], RAxML-NG 0.9.0[25]. The ANN was implemented in PyTorch[42] version 1.13.1.

# End notes

## Acknowledgements

We acknowledge the Data Science & AI Center at TAU for supporting this study. D.A. was supported by The Council for Higher Education program for excellent Ph.D. students in Data Sciences and by a fellowship from the Fast and Direct Ph.D. Program at Tel Aviv University. M.A. was supported by a fellowship from the Edmond J. Safra Center for Bioinformatics at Tel Aviv University. O.G. and Y.M. were supported by a grant from the European Research Council (ERC) under the European Union's Horizon 2020 research and innovation program (grant agreement No. 882396), by the Israel Science Foundation (grant number 993/17), the Yandex Initiative for Machine Learning at Tel Aviv University and a grant from the Tel Aviv University Center for AI and Data Science (TAD). T.P. was supported by an Israel Science Foundation grant 2818/21. I.M. was supported by an Israel Science Foundation grant 1843/21.

## Author Contributions

D.A., O.G., and M.A. jointly designed and conducted the work including programming the algorithm, performing the analyses, and drafting the manuscript. Y.M., T.P., and I.M. supervised this work and revised the manuscript.

## Competing interests

The authors declare no competing interests.

Supplementary Information

# The tree reconstruction game: phylogenetic reconstruction using reinforcement learning

Azouri, Granit, Alburquerque et al.

**This PDF file includes:**

Supplementary Figures 1-3
Supplementary Tables 1
Supplementary information

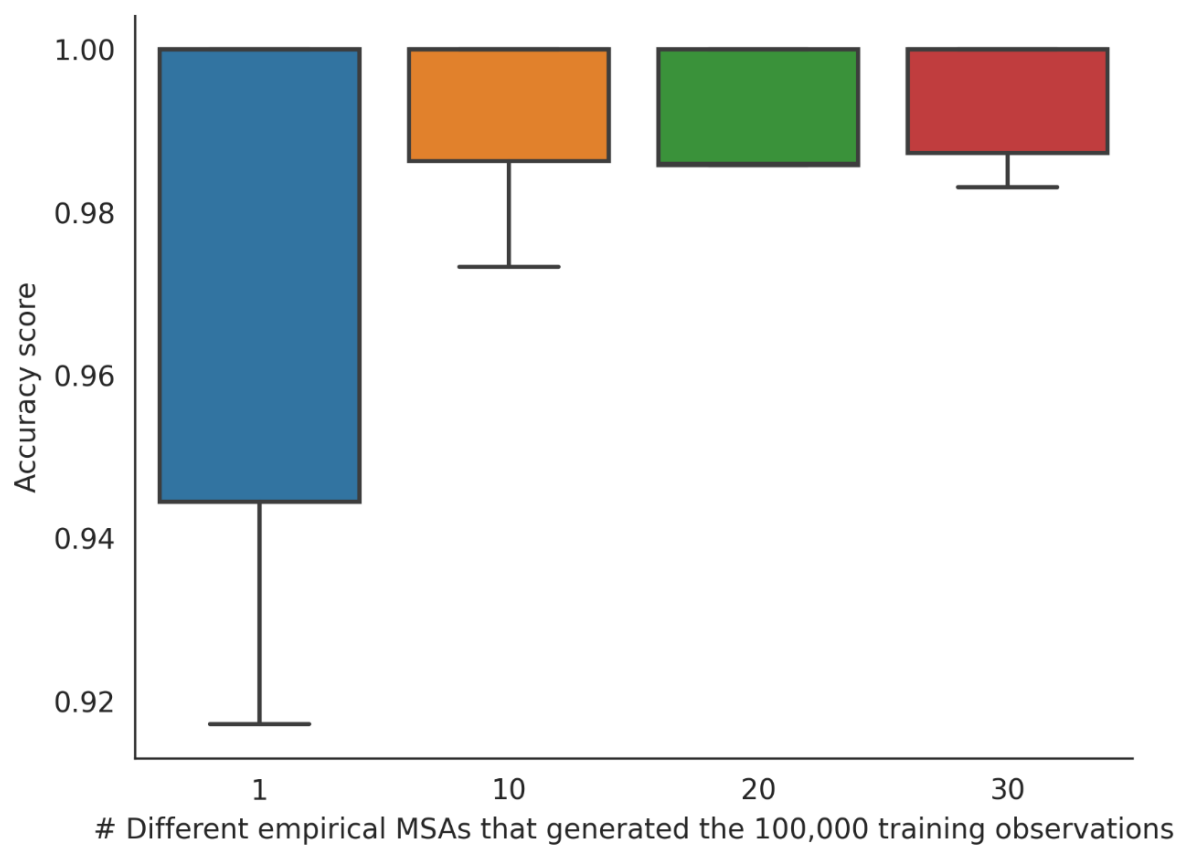

**Supplementary Figure 1. The impact of the number of datasets used to generate the training data on the prediction accuracy**. The RL model performance on the validation set containing 12 sequences (y axis) when using an increasing number of distinct datasets to generate the transition data. The box shows the quartiles of the dataset while the whiskers extend to show the $1.5 \times$ IQR past the low and high quartiles.

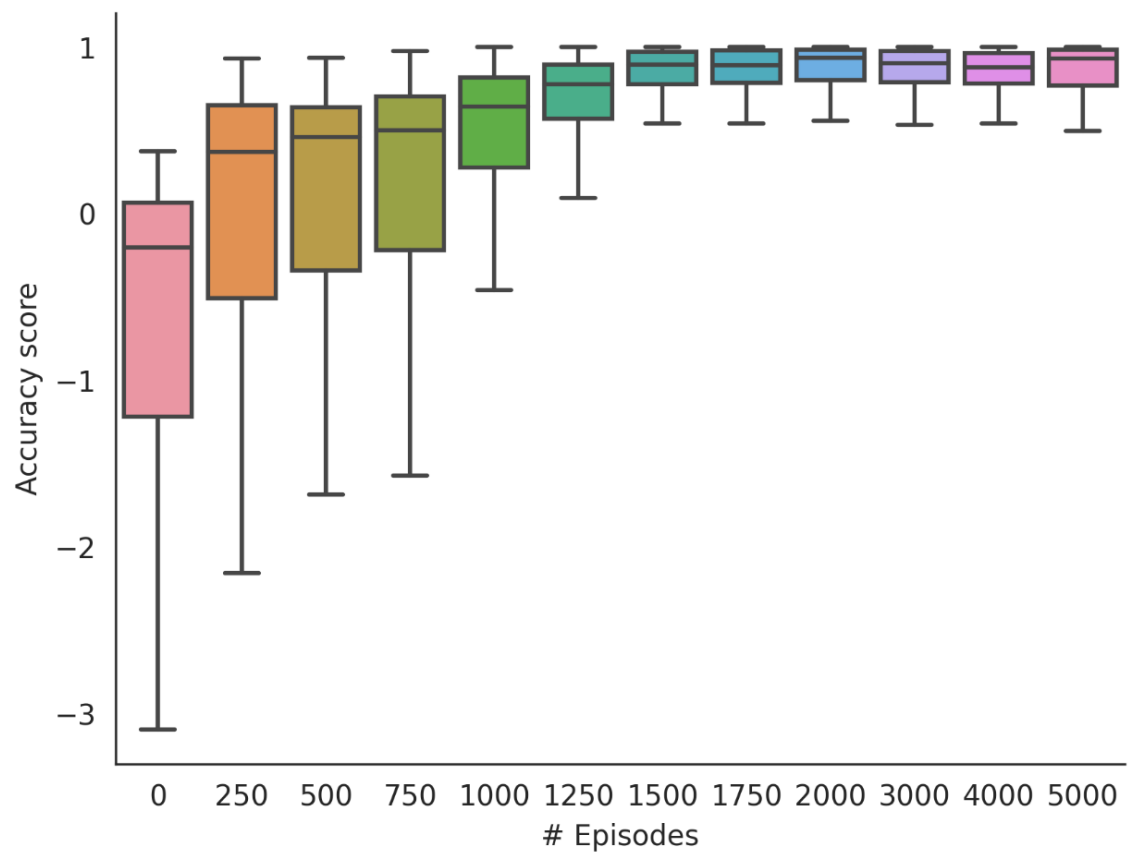

**Supplementary Figure 2. The impact of the number of training episodes on the validation accuracy.** The accuracy performance on 10 distinct empirical validation datasets (y axis) when using an increasing number of episodes in the training phase (x axis). The box shows the quartiles of the dataset while the whiskers extend to show the 1.5 × IQR past the low and high quartiles.

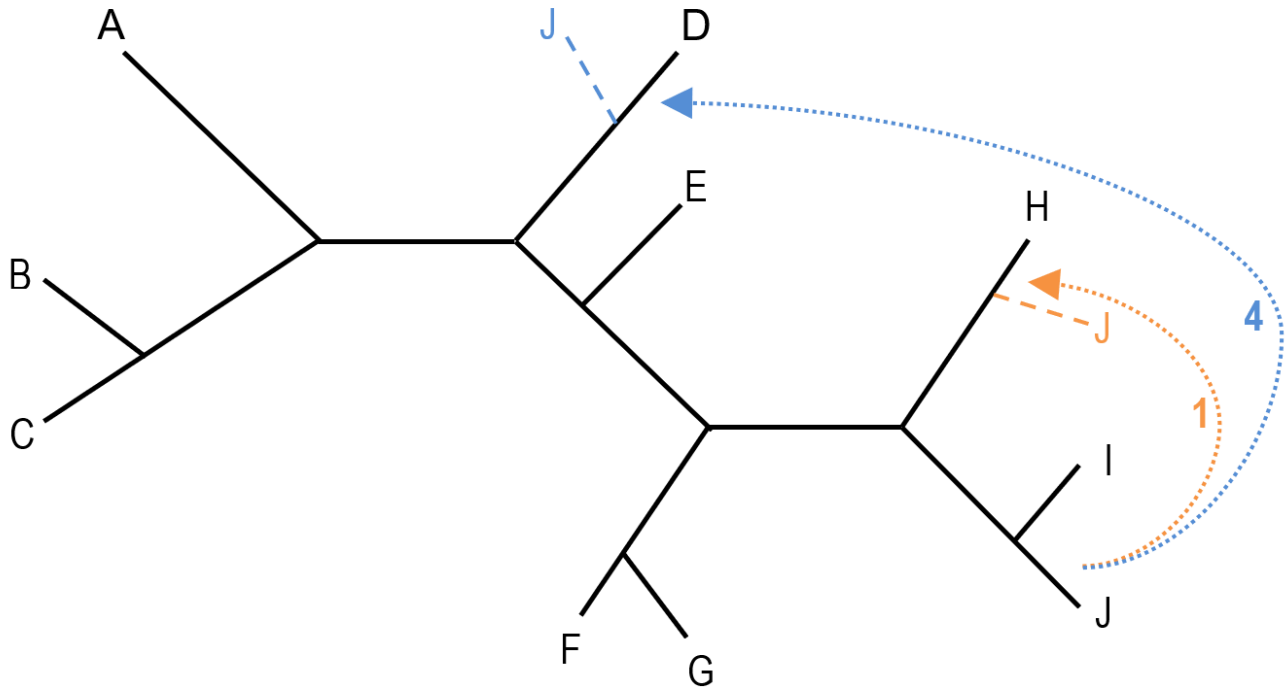

**Supplementary Figure 3. The radius restriction on SPR moves for relatively large datasets.** An illustrated example of an SPR move of radius one ("1"; in orange) and of radius four ("4"; in blue). In the developed RL framework, this restriction was applied for datasets containing 15 and 20 sequences.

**Supplementary Table 1.** The table details the hyperparameters values and further details of the RL configuration.

| Parameter name | Value in the trained model | Additional details |
| --- | --- | --- |
| NN architecture | Five fully connected hidden layers, in addition to the input layer (containing 27 neurons) and the output layer (containing a single node) | Number of neurons within each layer: {1: 4096, 2: 4096, 3: 2048, 4: 128, 5: 32} |
| Loss function | Mean Square Error | |
| Activation function | Leaky ReLU[52,53] | |
| Optimizer | Adam[54] | |
| Discount factor ($\gamma$) | 0.9 | |
| Replay buffer size | 10,000 | The maximal size of transitions collected during training |
| Times-to-learn | 50 | The number of times we sampled a batch to train the ANN |
| Horizon $H$ | 20 (for data of 7, 12, and 15 sequences), 30 (for 20 sequences) | The number of SPR moves in each episode. This hyperparameter was reoptimized when we considered different number of sequences in the analysis |
| Batch size | 128 | |
| Learning rate | $10^{-5}$ | |
| Exploratory policy | SoftMax | With T parameter = 1 |
| Episodes | 2,000 | Number of episodes in training |

## Supplementary Note 1: Optimizing the $Q$ function

In reinforcement learning, the Q-value function is a measure of the expected return (i.e., the immediate reward and future discounted rewards) of a particular action in a particular state. It is denoted as $Q(s, a)$, where $s$ is the state and $a$ is the action. When an agent is in a state $s$, it will choose the action with the highest Q-value, i.e., the action that is expected to yield the highest return. The Q-value function is typically estimated through trial and error, using an algorithm such as Q-learning. As the agent interacts with the environment, it updates its estimates of the Q-values based on the rewards it receives. Over time, the Q-value function converges to the optimal value, enabling the agent to make the best decisions possible in any given state. The optimal Q-value function, denoted as $Q^*$, represents the expected return of the best possible action in any given state. It can be used to determine the optimal policy for an agent, which is the set of rules that maximizes the agent's cumulative reward in an environment. In our case, $Q(s, a)$ corresponds to the maximal $\mathbb{E}_{s_0=s,\ a_0=a} [\sum_{t=0}^{H-1} \gamma^t r_t(s_t, a_t)]$, which we assume to be close to $Q^*(s, a)$. Notice that for any given dataset, there exists a $\gamma = 1 - \epsilon$, $\epsilon > 0$, which induces a $Q^*$ such that the optimal policy leads to the optimal topology from any starting topology. As an example, suppose we are in an environment with five states, $\{s_0, s_1, \ldots s_4\}$ where $s_2, s_4$ are states, and the reward for transitioning from $s_0$ to $s_1$ is $\delta$, from $s_0$ to $s_3$ is 1, from $s_1$ to $s_2$ is 1, and from $s_3$ to $s_4$ is 0:

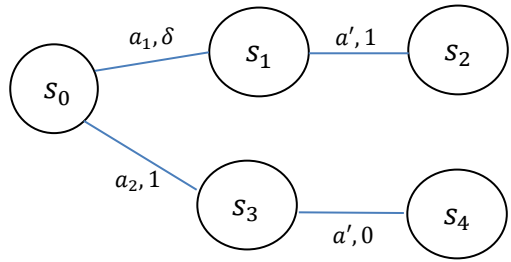

$$Q^*(s_0, a_1) = \delta + \gamma \cdot 1$$

$$Q^*(s_0, a_2) = 1 + \gamma \cdot 0 = 1$$

Here, $s_2$ is the optimal terminal state, but under a certain $\delta$ value the $Q$ value is greater for $s_4$. specifically, under $\gamma < 1 - \delta$ the optimal policy in $s_0$ is $a_2$, which leads us away from the global optimal topology $s_2$.

## Supplementary Note 2: Data collection for 20 sequences

The data collection strategy described throughout the paper was applied when training agents for 7, 12, 15, and 20 sequences. For datasets containing 20 sequences, we tested an additional strategy, in which we trained several agents, each on a single MSA, such that each constructed a different memory buffer. Next, all the memory buffers were combined to a single, larger, buffer of training observations. A new agent was then trained on this combined buffer. The purpose of training several agents on each single MSA was to simplify the RL environment, enabling our agents to collect observations with high-likelihood. Empirically, this agent achieved an average test accuracy of 0.89, while a regular agent (trained as described in the main text for smaller datasets) achieved an average test accuracy of 0.84 on MSAs containing 20 sequences.